\documentclass[12pt]{article}
\usepackage{amsfonts}
\usepackage{amsmath}
\usepackage{amsbsy}
\usepackage{amssymb}
\begin{document}
\hskip-20ex

\begin{center}
{\bf APPLICATION OF MONODROMY TRANSFORM METHOD TO SOLUTION\\
OF EINSTEIN - MAXWELL EQUATIONS WITH ISOMETRIES}\\[1ex]
G.A.~Alekseev\\[1ex]
V.A.~Steklov Mathematical Institute, AN SSSR, Moscow
\end{center}

\begin{abstract}
\noindent Applications of the monodromy transform approach to construction of exact solutions of electrovacuum Einstein - Maxwell field equations are considered. Examples of new solutions are given.
\noindent
\end{abstract}

The suggested earlier approach to solution of Einstein - Maxwell equations for electrovacuum fields depending on two coordinates [1,2], which was called as monodromy transform method, allows to reduce the problem to solution of one scalar singular integral equation. This integral equation looks as possessing a standard classical form [3,4]:
\begin{equation}\label{1}
\displaystyle\int\limits_L\dfrac{\mathcal{K}(\zeta,\tau)}{\zeta-\tau}
\,\mathbf{x}(\zeta)\,d\zeta=
\mathbf{k}(\tau)
\end{equation}
where the integral is understood as the Cauchy principal value integral on the cut $L$ on the spectral plane; $\tau$ and $\zeta\in L$. The kernel $\mathcal{K}(\zeta,\tau)$ and the right hand side
$\mathbf{k}(\tau)$ are given and $\mathbf{x}(\zeta)$ is unknown function. The entire structure of the field equations is "encoded" in the functions $\mathcal{K}(\zeta,\tau)$ and $\mathbf{k}(\tau)$
and in the structure of the cut $L$.

The cut $L$ on the complex plane $\tau$ should consist of two nonintersecting cuts: $L=L_+\cup L_-$, where $L_+$ joins $\tau=\xi_o$ and $\tau=\xi$, while $L_-$ joins $\tau=\eta_o$ and $\tau=\eta$. Here the parameters $\xi$ and $\eta$ are two real coordinates of the form: $\xi=x-t$, $\eta=x+t$ for wave fields or complex conjugated to each other, e.g., for stationary axisymmetric fields: $\xi=z+i \rho$, $\eta=z-i \rho$. The parameters $\xi_o$ and $\eta_o$ are the coordinates of some chosen fixed point at which the components of the unknown metric are normalized by the values of the components of Minkowski metric.

The expression for the kernel $\mathcal{K}(\zeta,\tau)$ includes the jump at the point $\zeta$ on the cut $L$ of the function $\lambda(\tau,\xi,\eta,\xi_o,\eta_o)$ and the scalar product of two vectors $\mathbf{k}(\tau)$ and $\mathbf{l}(\zeta)$:
\[\begin{array}{lcl}
\mathcal{K}(\zeta,\tau)=-\left[\lambda\right]_\zeta\, \left(\mathbf{k}(\tau)\cdot\mathbf{l}(\zeta)\right),
&&\mathbf{k}(\tau)=\left\{1,\,\mathbf{u}(\tau),\, \mathbf{v}(\tau)\right\},\\[1ex]
\lambda=\left[\dfrac{(\tau-\xi)(\tau-\eta)} {(\tau-\xi_o)(\tau-\eta_o)}\right]^{1/2},&&
\mathbf{l}(\zeta)=\begin{pmatrix}
1-i\epsilon(\zeta-\beta_o)\mathbf{u}^\dagger(\zeta)\\
i\epsilon(\zeta-\beta_o)+ \epsilon \alpha_o^2 \mathbf{u}^\dagger(\zeta)\\
-4\epsilon(\zeta-\xi)(\zeta-\eta)\mathbf{v}^\dagger(\zeta)
\end{pmatrix}.
\end{array}
\]
Here $\epsilon=1$ for real $\xi$,$\eta$ and $\epsilon=-1$ for complex conjugated ones; $\beta_o=(\xi_o+\eta_o)/2$, $\alpha_o=(\xi_o-\eta_o)/2j$, with $j=1$ for $\epsilon=1$ and $j=i$ for $\epsilon=-1$; $\mathbf{u}^\dagger(\tau)=\overline{\mathbf{u}(\overline{\tau})}$, $\mathbf{v}^\dagger(\tau)=\overline{\mathbf{v}(\overline{\tau})}$. The functions $\mathbf{u}(\tau)$ and  $\mathbf{v}(\tau)$ are arbitrary holomorphic on $L$ and therefore, we have actually not two, but four arbitrary functions: $\mathbf{u}_+(\tau)$ and $\mathbf{v}_+(\tau)$ should be holomorphic on $L_+$, while $\mathbf{u}_-(\tau)$ and $\mathbf{v}_-(\tau)$ should be holomorphic on $L_-$. For vacuum fields $\mathbf{v}_{\pm}(\tau)\equiv 0$ and there remains only two arbitrary functions $\mathbf{u}_{\pm}(\tau)$.

The right hand side of the scalar equation (\ref{1}) is one of the components of the vector $\mathbf{k}(\tau)=\left\{1,\, \mathbf{u}(\tau),\, \mathbf{v}(\tau)\right\}$. Let  $\mathbf{x}_{[u]}$ and $\mathbf{x}_{[v]}$ are the solutions of (\ref{1}) corresponding to the choice in the right hand side of (\ref{1}) the functions $\mathbf{u}(\tau)$ and $\mathbf{v}(\tau)$ respectively. Then the Ernst potentials possess the expressions
\[\mathcal{E}=-\epsilon-\dfrac 2{\pi}\displaystyle{\int\limits_L} [\lambda]_\zeta\, f(\zeta)\,\mathbf{x}_{[u]}(\zeta)\, d\zeta,\qquad
\Phi=\dfrac 2{\pi}\displaystyle{\int\limits_L} [\lambda]_\zeta\, f(\zeta)\,\mathbf{x}_{[v]}(\zeta)\, d\zeta
\]
where $\epsilon=\pm1$ and the function $f(\zeta)=1-i\epsilon(\zeta-\beta_o)\mathbf{u}^\dagger(\zeta)$.

Various applications of the method described above, the properties of the functions $\mathbf{u}(\tau)$, $\mathbf{v}(\tau)$ and their relation to the propeties of the corresponding solutions have been considered in [2],[5]. Here it is important to recall that if $\mathbf{u}_+$ and $\mathbf{u}_-$ and similarly, $\mathbf{v}_+$ and $\mathbf{v}_-$ are analytically continuations of each other and are arbitrary rational functions, then the corresponding solution of the field equations possess the explicit expressions in elementary functions [2] which can be obtained using the elementary theory of residues. This allows to construct numerous examples of new exact solutions and to extend considerably the known classes of solutions by including there new parameters as well as to construct the nonlinear superpositions of the fields of various sources or waves.

Below we present some examples corresponding to the simplest choices of rational functions $\mathbf{u}(\tau)$ and $\mathbf{v}(\tau)$ in which we restrict our consideration by the stationary electrovacuum fields with axial symmetry.
\[\begin{array}{l}
\begin{array}{lcl}
\begin{array}{l}\text{1)}\end{array}\hskip1ex \mathbf{u}(\tau)=0, \mathbf{v}(\tau)=0&\text{--}&
\begin{array}{l}
\text{Minkowski space-time}
\end{array}\\[1ex]
\begin{array}{l}\text{2)}\\{} \end{array}
\hskip-0.5ex\left\{\hskip-0.5ex\begin{array}{l}
\mathbf{u}(\tau)=u_o=const\\
\mathbf{v}(\tau)=0
\end{array}\right. &
\begin{array}{c}
\text{--}\\
{}
\end{array}&
\begin{array}{l}
\text{Minkowski space-time in accelerated reference frame}\\
\text{(Rindler metric)}
\end{array}\\[2ex]
\begin{array}{l}\text{3)}\\{} \end{array}
\hskip-0.5ex\left\{\hskip-0.5ex\begin{array}{l}
\mathbf{u}(\tau)=u_o=const\\
\mathbf{v}(\tau)=v_o
\end{array}\right. &
\begin{array}{c}
\text{--}\\
{}
\end{array}&
\begin{array}{l}
\text{Space-time filled by a homogeneous electromagnetic field}\\
\text{(Bertotti-Robinson universe)}
\end{array}\\[2ex]
\begin{array}{l}\text{4)}\\{} \end{array}
\hskip-0.5ex\left\{\hskip-0.5ex\begin{array}{l}
\mathbf{u}(\tau)=a\,\tau+b\\
\mathbf{v}(\tau)=c\,\tau+d
\end{array}\right. &
\begin{array}{c}
\text{--}\\
{}
\end{array}&
\begin{array}{l}
\text{Generalization of Melvin electromagnetic universe}\\
\text{(including the Bertotti-Robinson solution)}
\end{array}
\end{array}
\\[10ex]
\hskip2ex\text{5)}\hskip3ex \mathbf{u}(\tau)=\dfrac{u_1}{\tau-h_1},\,\, \mathbf{v}(\tau)=\dfrac{v_1}{\tau-h_1}\quad\text{--}\quad
\text{a complete family of Kerr-Newman solutions}\\[2ex]
\hskip2ex\text{6)}\hskip3ex \mathbf{u}(\tau)=\dfrac{u_1}{\tau-h_1}+\dfrac{u_2}{\tau-h_2},\,\, \mathbf{v}(\tau)=\dfrac{v_1}{\tau-h_1}+\dfrac{v_2}{\tau-h_2}
\quad\text{--}\quad
\text{the field of two interacting sources}\\[-1ex]
\hskip62ex\text{of the Kerr-Newman type.}
\end{array}
\]

For a conclusion we present three examples of solutions for static fields which are the particular cases of 6) written in terms of bipolar coordinates centered at the sources:
\[\begin{array}{l}
1)\quad g_{tt}=\dfrac{(r_1^2-2 m_1 r_1+e_1^2)(r_1^2-2 m_1 r_1 + e_1^2)}{(r_1 r_2-e_1 e_2)^2},\qquad \Phi=\dfrac{e_1(r_2-m_2)+e_2(r_1-m_1)}{r_1 r_2-e_1 e_2},\qquad \dfrac{e_1}{m_1}=\dfrac{e_2}{m_2}.\\[2ex]
2)\quad g_{tt}=\left\{\dfrac{\bigl[\ell^2-m_1^2-m_2^2+2 m_1 m_2\cos(\theta_1-\theta_2)\bigr](r_1-m_1)(r_2-m_2)} {\bigl[\ell^2-(m_1-m_2)^2\bigr] r_1 r_2+m_1 m_2(r_1-r_2)^2}\right\}^2,\\[4ex]
\phantom{2)\quad }\Phi=\dfrac{\sqrt{(\ell^2-m_1^2-m_2^2)^2-4 m_1^2 m_2^2}(m_1 r_2-m_2 r_1)}{\bigl[\ell^2-(m_1-m_2)^2\bigr] r_1 r_2+m_1 m_2(r_1-r_2)^2},\qquad\qquad
\begin{array}{l} e_1=m_1,\quad e_2=-m_2,\\
\ell=z_1-z_2.\end{array}\\[3ex]
3)\quad g_{tt}=\dfrac{(r_1^2+d^2)(r_2^2+d^2)}{(r_1 r_2+d^2)^2},\qquad
\Phi=i\dfrac{d(r_2-r_1)}{r_1 r_2+d^2}.
\end{array}
\]
Here $g_{tt}=\mathcal{E}+\Phi \overline{\Phi}$. In the first example which is known, but which seems as has not been presented in the literature in a so simple form, the charges are proportional to masses; in the second one -- the signs of charges are different but their absolute values are equal to masses; in the third case, we have the simplest massless magnetic dipole.

{\small
\begin{list}{}{}
\item{\hskip-4.5ex [1]\hskip2ex G.A.~Alekseev, DAN SSSR, v.283, no. 3, pp. 577 - 582 (1985); English transl. Sov.Phys.Dokl., {\bf 30}, pp. 565--568 (1985).}
\vspace{-2ex}
\item{\hskip-4.5ex [2]\hskip2ex G.A.~Alekseev, Trudy MIAN SSSR, v.176, no. 4, pp. 211 - 258 (1987);  English transl.  Proc. Steklov Math. Inst., Providence, RI: American Math. Soc., {\bf 3}, pp. 215 -- 262 (1988).}
\vspace{-2ex}
\item{\hskip-4.5ex [3]\hskip2ex N.I.~Muskhelishvili, Singular integral equations, 3rd ed.,"Nauka", Moscow, 1968; English transl. of 1st ed., Noordhoff, 1953; reprinted, (1972).}
\vspace{-2ex}
\item{\hskip-4.5ex [4]\hskip2ex F.D.~Gakhov, Boundary value problems, 3rd ed.,"Nauka", Moscow, 1977; English transl. of 2nd ed., Pergamon Press, Oxford and Addison-Wesley, Reading, Mass., (1966).}
\vspace{-2ex}
\item{\hskip-4.5ex [5]\hskip2ex G.A.~Alekseev, Proceedings of the 2nd All-Union Workshop "Exact solutions of gravitational field equations and their physical interpretation", Tartu, pp. 6--9 (1988) (in Russian).}
\end{list}
}
\end{document}